\documentclass[a4paper,11pt]{article}
\pdfoutput=1

\usepackage{jcappub}
\usepackage{subcaption}
\usepackage{xcolor}
\usepackage{braket}
\usepackage{soul}
\usepackage{slashed}

\title{Constraining axion quadratic couplings with the Hulse-Taylor binary system}

\author{Ziwen Yin$^{1,2}$}
\affiliation{$^1$ Tsung-Dao Lee Institute (TDLI),
No.\ 1 Lisuo Road, 
 201210 Shanghai, China;}
\affiliation{$^2$ School of Physics and Astronomy, Shanghai Jiao Tong University,
800 Dongchuan Road, 200240 Shanghai, China}
\author{Shyam Balaji$^3$}
\author{Malcolm Fairbairn$^3$}
\author{David J. E. Marsh$^3$}
\affiliation{$^3$Theoretical Particle Physics and Cosmology, Department of Physics, King's College London, Strand, London, WC2R 2LS, United Kingdom}

\emailAdd{ziwenyin@sjtu.edu.cn, david.j.marsh@kcl.ac.uk}

\abstract{The orbital evolution of the Hulse-Taylor binary neutron star system is described to high precision by general relativity, in which gravity is the only long-range force and gravitational waves provide the dominant energy-loss channel. We use this precision test of relativistic binary dynamics to derive new constraints on axion couplings to stable neutron star constituents: neutrons, electrons, and muons. Quadratic shift symmetry breaking axion-fermion couplings allow binary systems to lose energy through dipole and quadrupole emission of axion waves. These couplings also mediate long range, spin independent forces in two different regimes: in an ambient dark matter  background, and when a tachyonic phase transition is triggered inside the neutron stars. For light QCD axions our constraints can be recast as limits on the axion decay constant, which are complementary to other probes for $m_a\lesssim 10^{-12}\text{ eV}$ and $f_a\lesssim M_\textrm{pl}$. For $m_a\lesssim10^{-12}\,{\rm eV}$, we place, to our knowledge, the strongest available constraint on the quadratic axion--muon coupling scale, providing a complementary probe to supernova cooling.}


\begin{document}
\maketitle
\flushbottom

\section{Introduction}
\label{sec:introduction}

The orbital decay of the Hulse-Taylor binary system, observed now for over 50 years, provided the first indirect evidence for the existence of gravitational waves (GWs), and combined with the orbit itself provides a powerful test of general relativity and gravity as the dominant force governing the evolution of the system~\cite{1982ApJ...253..908T}. Light scalar fields can mediate long-rangeforces and provide new orbital energy loss channels via scalar waves, and so the Hulse-Taylor system can be used to set robust limits on the possible existence and coupling strength of particles beyond the Standard Model~\cite{Mohanty:1994yi}.

A particularly well-motivated class of new (pseudo) scalar fields are axions (see reviews Refs.~\cite{Marsh:2015xka,Chadha-Day:2021szb,OHare:2024nmr}). Axions can solve the strong-CP problem~\cite{Peccei:1977hh,Weinberg:1977ma,Wilczek:1977pj} caused by the anomalously small neutron electric dipole moment~\cite{Abel:2020gbr}, provide a well-motivated candidate for the observed~\cite{Planck:2018vyg} dark matter (DM)~\cite{Marsh:2024ury}, and are a generic prediction of string theory compactifications~\cite{Svrcek:2006yi,Arvanitaki:2009fg,Cicoli:2012sz,Demirtas:2021gsq}. Axion DM is typically produced non-thermally in the early universe by the vacuum realignment mechanism~\cite{Preskill:1982cy,Dine:1982ah,Abbott:1982af}, which allows for very low particle masses, $m_a\ll 1\text{ eV}$, while still behaving as cold DM for $m_a>2.2\times 10^{-21}\text{ eV}$~\cite{Zimmermann:2024xvd}. The solution to the strong-CP problem demands that the QCD axion couples to nucleons, while symmetry also allows other axions to couple directly to fermions. Axions can thus mediate new macroscopic forces~\cite{Moody:1984ba}, potentially over astronomical distances.

Existing searches for axion-fermion interactions probe several qualitatively distinct classes of operators. The derivative axial-vector coupling is shift invariant and CP preserving, and gives rise to spin-dependent forces~\cite{Moody:1984ba} and direct-detection signals (see e.g. Refs.~\cite{Graham:2013gfa,Garcon:2017ixh,Garcon:2019inh} and references therein). By contrast non-derivative, linear, scalar or pseudoscalar couplings break the perturbative shift symmetry and can generate CP-violating monopole-dipole or monopole-monopole forces~\cite{Moody:1984ba,Arvanitaki:2014dfa}. The quadratic couplings considered in this work are CP preserving and spin-independent. For a purely shift-invariant axion, their leading effects can cancel or be related to higher-order terms in the derivative interaction, whereas shift-symmetry-breaking contributions can survive as genuine long-range scalar charges~\cite{Bauer:2023czj,Kim:2023pvt}. Laboratory, astrophysical, and cosmological probes therefore constrain complementary regions of parameter space: torsion-balance and precision-force experiments are especially powerful for macroscopic forces~\cite{Adelberger:2003zx,Arvanitaki:2014dfa,Bauer:2023czj}, stellar and supernova cooling constrain light particles coupled to electrons, nucleons, and muons~\cite{Raffelt:1999tx,Hook:2017psm,Croon:2020lrf,Balaji:2022noj,Day:2023mkb,DelaTorreLuque:2023huu,DelaTorreLuque:2024zsr,Fiorillo:2026wso,Caputo:2021rux,Bottaro:2023gep,Fiorillo:2025sln}, and recent work has emphasized that an ambient axion-DM background can enhance forces from quadratic couplings and open new detection channels~\cite{Grossman:2025cov,Cheng:2025fak,VanTilburg:2024xib,Gan:2025nlu,Bouley:2026frx}.

In this work we focus on the dimension-five quadratic coupling
\begin{align}
    \mathcal{L}_{\rm int} = \frac{a^2}{\Lambda_\psi}\bar{\psi}\psi\, ,
\end{align}
where $\Lambda_\psi$ is the effective scale to be constrained and $\psi$ denotes a neutron star constituent, in particular a nucleon, electron, or muon. This operator generates a spin-independent interaction between macroscopic objects. In an ambient axion-DM background, the quadratic coupling induces an effectively long-range force whose strength can be enhanced by the coherent axion field~\cite{Grossman:2025cov,Cheng:2025fak,VanTilburg:2024xib,Gan:2025nlu}, or lead to a tachyonic phase transition~\cite{Hook:2017psm,Huang:2018pbu}. The same coupling also acts as a source for dipole and quadrupole axion radiation from binary motion. Assuming a ``light QCD axion'' constitutes DM, the matter effect introduced by the axion-matter quadratic coupling inside compact objects will reduce the effective mass of the axion. This leads compact objects (e.g. the earth and neutron stars) to behave like a material with refractive index larger than one for the axion DM wind. This effect can enhance the axion field gradient near the interface of the object interior and exterior, which may benefit searches for the axion gradient couplings in experiments on the earth~\cite{Banerjee:2025dlo}.

Neutron stars are especially useful laboratories because they contain large numbers of nucleons and, through beta-equilibrium, stable electron and muon populations. Typical neutron star equations-of-state allow lepton fractions $Y_e\sim Y_\mu\sim\mathcal{O}(10^{-2})$, making muon- and electron-coupled forces observable in principle through binary dynamics~\cite{cohen1970neutron,PhysRevD.100.035039,Zhang:2020wov,Dror:2019uea}. The Hulse--Taylor binary pulsar therefore constrains these operators through both fifth-force effects and additional energy loss. Phenomenologically, the force has two regimes relevant for this work: a background-enhanced interaction in the axion-DM background~\cite{Grossman:2025cov,Cheng:2025fak,VanTilburg:2024xib,Gan:2025nlu, Bouley:2026frx}, and a force sourced by a tachyonic axion profile inside neutron stars~\cite{Hook:2017psm,Huang:2018pbu}.

This paper is organised as follows. In Sec.~\ref{sec:gr_tests} we review the Hulse--Taylor observables used to test deviations from general relativity and introduce the phenomenological parameters describing fifth forces and additional radiation. In Sec.~\ref{sec:axion_forces} we describe the relevant axion-fermion couplings, the background-enhanced force, the tachyonic neutron star profile, and axion radiation from binary motion. In Sec.~\ref{sec:results} we present the resulting constraints on quadratic axion couplings to neutron star constituents. We conclude in Sec.~\ref{sec:conclusions}. Appendix~\ref{sec:HTGR} provides a detailed and pedagogic description of how binary systems test general relativity and constrain new forces. Appendix~\ref{sec:mumu_annihilation} summarizes supernova cooling estimates for quadratic axion-muon and axion-electron couplings.

\section{Statistical analysis of the Hulse--Taylor system}
\label{sec:gr_tests}

The Hulse--Taylor binary pulsar PSR B1913+16 provides three precisely measured post-Keplerian observables: the periastron advance $\dot{\omega}$, the Einstein delay parameter $\gamma$, and the intrinsic orbital-period derivative $\dot{P}_b$, see Table.~\ref{tab:hulse_taylor_inputs}.~\footnote{Our Table corrects a typo: the value for $\gamma$ in Ref.~\cite{Weisberg:2016jye} Table 2 is quoted with incorrect units.} 
In the presence of an additional long-range force and exotic radiation channels, these observables depend on the phenomenological parameters $\beta$ and $\xi$, which parameterise modifications to the conservative dynamics and the total radiative energy loss, respectively, such that
\[
V(r)=-\frac{G(1+\beta)M_1M_2}{r},
\qquad
\dot E_{\rm rad}=(1+\xi)\dot E_{\rm GW}^{\rm GR},
\]
where $G$ is Newton's constant. The sign convention is therefore such that $\beta>0$ is an additional attractive inverse-square force, $\beta<0$ is effectively repulsive/weaker attraction, and $\xi>0$ denotes additional positive energy loss due to scalar radiation leaving the system. 
\begin{table}[!t]
\centering
\begin{tabular}{lll}
\hline
Parameter & Value & Comment \\
\hline
$P_b$ 
& $0.322997448918\pm 0.000000000003~{\rm d}$ 
& Orbital period \\

$e$ 
& $0.6171340 \pm 0.0000004$ 
& Orbital eccentricity \\

$\dot{\omega}_{\rm obs}$ 
& $(4.226585 \pm 0.000004)~{\rm deg~yr^{-1}}$ 
& Periastron advance \\

$\gamma_{\rm obs}$ 
& $(4.307 \pm 0.004)\times 10^{-3}~{\rm s}$ 
& Einstein delay parameter \\

$\dot{P}_{b,{\rm obs}}$ 
& $(-2.423 \pm 0.001)\times 10^{-12}$ 
& Observed orbital-period derivative \\

$\dot{P}_{b,{\rm Gal}}$ 
& $(-0.025 \pm 0.009)\times 10^{-12}$ 
& Galactic correction \\

$\dot{P}_{b,{\rm int}}$ 
& $(-2.398 \pm 0.004)\times 10^{-12}$ 
& Intrinsic value used in the fit \\
\hline
\end{tabular}
\caption{
Timing parameters for PSR B1913+16 used in the $\chi^2$ analysis. 
The orbital period, eccentricity, periastron advance, Einstein delay parameter and orbital-period derivative are taken from Ref.~\cite{Weisberg:2016jye}. The intrinsic period derivative used for deriving constraints is obtained by subtracting the Galactic correction, $\dot{P}_{b,{\rm int}}=\dot{P}_{b,{\rm obs}}-\dot{P}_{b,{\rm Gal}}$.
}
\label{tab:hulse_taylor_inputs}
\end{table}
See Appendix.~\ref{sec:HTGR} for more details on how general relativity's predictions are modified in the presence of the axion field. For each point in the $(\beta,\xi)$ plane, we construct the chi-squared statistic
\begin{equation}
\chi^2(M_1,M_2;\beta,\xi)=
\left(
\frac{\dot{\omega}_{\rm th}-\dot{\omega}_{\rm obs}}
{\sigma_{\dot{\omega}}}
\right)^2
+
\left(
\frac{\gamma_{\rm th}-\gamma_{\rm obs}}
{\sigma_\gamma}
\right)^2
+
\left(
\frac{\dot{P}_{b,{\rm th}}-\dot{P}_{b,{\rm int}}}
{\sigma_{\dot P_{b,{\rm int}}}}
\right)^2.
\label{eq:chi2}
\end{equation}

The neutron star masses are treated as nuisance parameters and are profiled over by minimising Eq.~\eqref{eq:chi2},
\begin{equation}
\chi^2_{\rm prof}(\beta,\xi)=
\min_{M_1,M_2}
\chi^2(M_1,M_2;\beta,\xi),
\end{equation}
subject to the physically motivated range
\begin{equation}
1\,M_\odot \leq M_{1,2} \leq 2.5\,M_\odot.
\end{equation}

The global best-fit point corresponds to the minimum value $\chi^2_{\rm min}$, and exclusion regions are obtained from the profile likelihood through
\begin{equation}
\Delta\chi^2(\beta,\xi)=
\chi^2_{\rm prof}(\beta,\xi)-\chi^2_{\rm min}.
\end{equation}

For exclusion contours in the full $(\beta,\xi)$ plane, both parameters are treated as parameters of interest and we use
\begin{equation}
\Delta\chi^2 = 5.99 \quad (95\%~{\rm C.L.}),
\qquad
\Delta\chi^2 = 9.21 \quad (99\%~{\rm C.L.}).
\end{equation}

\begin{figure}[htbp]
    \centering
    \includegraphics[width=0.75\linewidth]{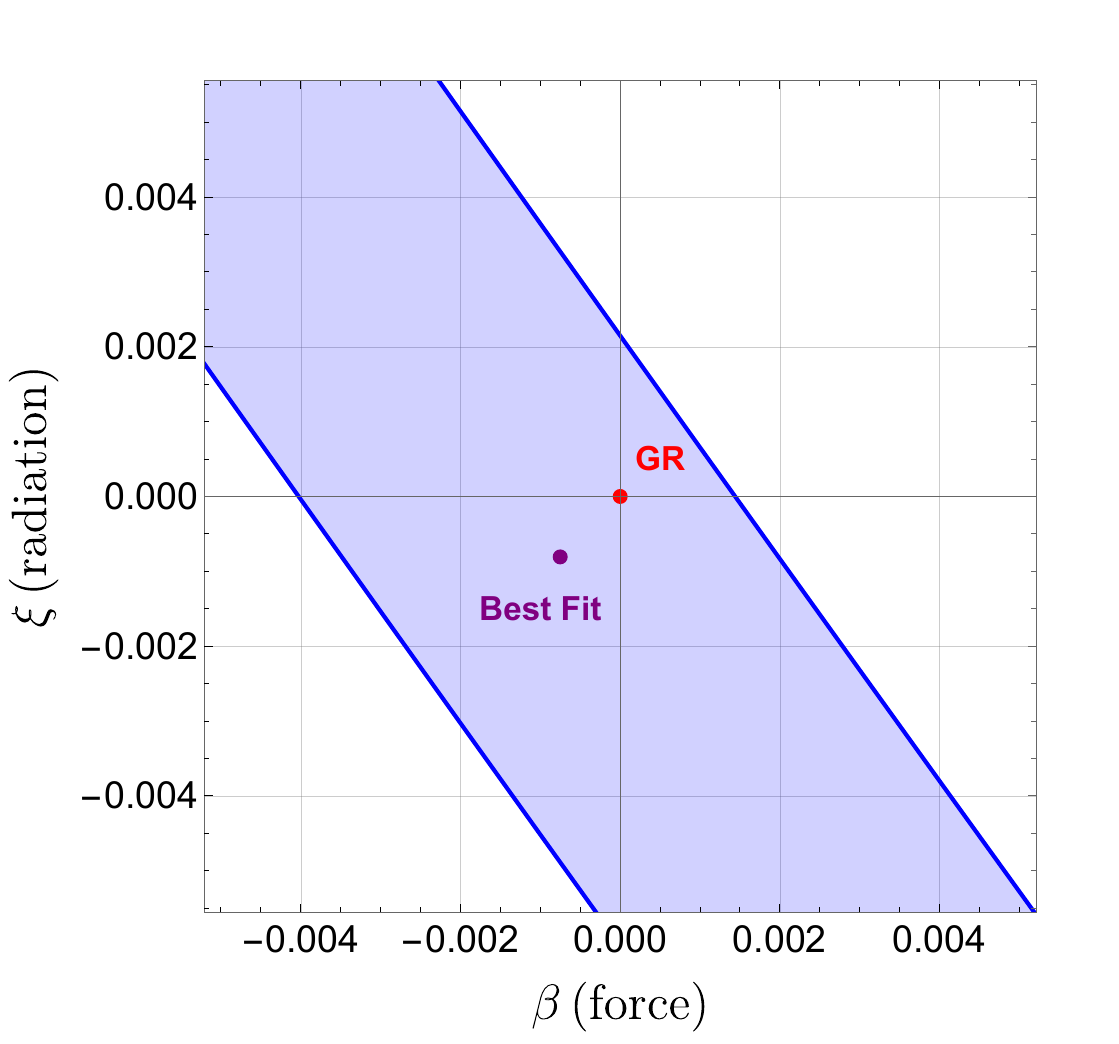}
    \caption{95\% CL region allowed by Hulse-Taylor measurements of PSR B1913 + 16 in the $\beta$-$\xi$ (force vs. radiation) parameter plane. General relativity corresponds to the origin shown in red with no additional effects beyond gravity, the ``Best Fit'' point corresponds to the global $\chi^2$ best fit point.}
    \label{fig:betaxi}
\end{figure}

The apparent \((1+\beta)^{3/2}\) scaling is not the fixed-mass scaling of the orbital-period derivative. At fixed neutron star masses, the modified quadrupole expression gives
\[
\dot P_b\propto (1+\xi)(1+\beta)^{2/3}.
\]
However, in the timing fit the masses are profiled over at each value of \(\beta\).  Since the periastron advance is measured very precisely, the fit approximately preserves
\[
\dot\omega_{\rm th}\propto M^{2/3}(1+\beta)^{-1/3},
\]
which implies \(M\propto (1+\beta)^{1/2}\). The Einstein delay parameter then approximately fixes the mass ratio, so that \(\mu\propto M\). The mass-dependent factor in the orbital-decay prediction therefore scales as
\[
\mu M^{2/3}\propto M^{5/3}\propto (1+\beta)^{5/6}.
\]
Combining this with the explicit conservative-force dependence in \(\dot P_b\) gives
\[
\dot P_{b,{\rm th}}
\propto
(1+\xi)(1+\beta)^{2/3}(1+\beta)^{5/6}
=
(1+\xi)(1+\beta)^{3/2}.
\]
Thus the degeneracy direction in the profiled \(\chi^2\) fit is approximately
\[
1+\xi \simeq C(1+\beta)^{-3/2},
\]
where \(C\) is fixed by the small offset between the observed intrinsic orbital decay and the general relativity prediction at the best-fit point. Hence we are left with the constraints

\begin{align}
    \xi-0.990015(1+\beta)^{-3/2}&>-0.996039 \\
    \xi-0.998121(1+\beta)^{-3/2}&<-0.995995\label{eq:T_dot}
\end{align}
or we can write it approximately as

\begin{equation}
0.993976<\frac{\Dot{P}_{\mathrm{b\,mod}}}{\Dot{P}_{b\,\mathrm{GR}}}=(1+\xi)(1+\beta)^{3/2}< 1.002126 \label{eq:T_dot}
\end{equation}

\section{Axion Forces and Axion Emission}
\label{sec:axion_forces}

The possible existence of light DM axions coupled to neutron star constituents modifies the evolution of binary systems, firstly by mediating a long-range force, and secondly by adding a new emission channel for energy loss.

\subsection{Axion-Fermion Couplings}

Axions are subject to a continuous perturbative shift symmetry, $a\rightarrow a+{\rm const.}$, which in field theory is inherited from a global U(1) symmetry and for closed string axions in string theory is inherited from gauge invariance of a higher form symmetry. Thus, the most natural coupling of axions to fermions is the shift symmetry invariant (SI) coupling
\begin{align}
    \mathcal{L}_{\rm int} = c_\psi \frac{(\partial_\mu a)}{2f_a}\bar{\psi}\gamma^\mu\gamma_5 \psi\, .
    \label{eqn:deriv_int}
\end{align}
Since axions are pseudoscalars, this coupling preserves CP symmetry. Performing a chiral rotation on the fermions and using the divergence of the axial vector current, this coupling can be rewritten in terms of the anomalous coupling of axions to gauge fields under which $\psi$ is charged, and non-derivative axion-fermion couplings. As discussed in e.g. Ref.~\cite{Bauer:2023czj}, the axion-fermion couplings have higher order contributions, in particular an SI quadratic coupling
\begin{align}
    \mathcal{L}^{(2)}_{\rm SI} = c_\psi^2  m_\psi \frac{a^2}{f_a^2}\bar{\psi}\psi\, .
    \label{eqn:L2_SI}
\end{align}

The axion continuous shift symmetry is generically broken by non-perturbative effects, which generate a mass for the axion. Shift symmetry breaking (SB) operators are thus also induced. Of particular interest is the SB quadratic fermion coupling
\begin{align}
    \mathcal{L}^{(2)}_{\rm SB} = \alpha_\psi \frac{a^2}{f_a^2}\bar{\psi}\psi\equiv \frac{a^2}{\Lambda_\psi}\bar{\psi}\psi\, .
      \label{eqn:L2_SB}
\end{align}
The scale of the dimensionful parameter $\alpha_\psi$ depends on the source of SB and the relation to the fermions $\psi$. In this work, we are interested in the cases where $\psi$ is a nucelon, electron or a muon, and place constraints on the scale $\Lambda_\psi$. If $\alpha_\psi$ is known assuming a particular model then we can place constraints directly on $f_a$.

If the axion obtains some of its mass from coupling to gluons and QCD instantons (like the QCD axion), and for $\psi=N$ (a nucleon), then $\alpha_N$ is related to the QCD scale, and thus to the nucleon mass. In particular for a coupling to gluons $c_{GG}$ and $m_a\ll m_\pi$ we have $\alpha_N\approx\frac{1}{2}\sigma_{N\pi}\frac{m_um_d}{(m_u+m_d)^2}$~\cite{Banerjee:2025dlo} or $\alpha_N \approx 5.4\text{ MeV}$. When considering $\psi=N$, in order to place constraints on $f_a$ we assume the axion is a ``light QCD axion'' (see e.g. Refs.~\cite{Bauer:2023czj,Hook:2017psm,Kim:2023pvt}), with $m_a=\epsilon \, m_{\rm QCD}$, with $m_{\rm QCD}$ the QCD axion mass~\cite{GrillidiCortona:2015jxo} (we note that such models with $\epsilon\ll 1$ are fine-tuned).

If the axion mass is the only parameter that breaks the shift symmetry, then one expects $\alpha_\psi = m_a^2/f_a\ll m_\psi$ for small values of $m_a$ and large values of $f_a$ of interest for axion DM. Quadratic SB operators for muons less suppressed than $m_a^2/f_a$ can be generated in quantum field theory by kinetic mixing with a confining dark sector~\cite{Beadle:2023flm}. For the electron coupling, Ref.~\cite{Kim:2023pvt} finds a two-loop induced coefficient $\alpha_e\simeq0.30~{\rm eV}$ for a KSVZ-like light QCD axion. The analogous muon coefficient is $\alpha_\mu=2.77\,\rm{eV}$. 
Less suppressed values of $\alpha_\psi$ can be generated for any axion (not just a light QCD axion), if the axion couples to an instanton that also leads to non-perturbative corrections to the fermion Yukawa couplings. Such a situation arises in string theory (e.g. Refs.~\cite{Blumenhagen:2006xt,Leontaris:2009ci,Cicoli:2012sz}). Expanding the relevant instanton effect, the constant piece generates an axion-independent correction to the fermion mass, $\delta m_\psi$, the linear piece generates a CP-violating axion-fermion coupling~\footnote{We neglect the CP-violating coupling in this work. The QCD axion in the Standard Model has small but calculable CP-violating couplings (see e.g. Ref.~\cite{Arvanitaki:2014dfa}). In string theory, following Refs.~\cite{Demirtas:2021gsq,Gendler:2023kjt}, CP violating effects for all but the very lightest axions become negligible in models with many axions.}, and the quadratic piece takes the form of Eq.~\eqref{eqn:L2_SB} with $\alpha_\psi\approx \delta m_\psi$. For muons, the scale of $\delta m_\mu$ could plausibly range from the experimental error on the muon mass, around 2 eV, down to the scale of neutrino masses, $10^{-3}\text { eV}$.

\subsection{Axion-Mediated Forces and Background Enhancement}

Classical long-range forces can be computed in quantum field theory according to the Born approximation which relates the amplitude, $\mathcal{M}$, to the momentum space potential, $\tilde{V}(q)$, in the non-relativistic limit. The real space potential is obtained via a Fourier transform. The classic computation of axion mediated forces by Ref.~\cite{Moody:1984ba} considers both CP-preserving and CP-violating couplings at linear order and evaluates the potential at tree-level. The pure CP-preserving force is generated by the first term in the pseudo-scalar basis after rotating Eq.~\eqref{eqn:deriv_int}
\begin{align}
    \mathcal{L}^{(1)}_{\rm SI} = -ic_\psi m_\psi \frac{a}{f_a}\bar{\psi}\gamma_5\psi\, .
\end{align}
This term generates a spin-dependent, dipole-dipole short-range force, which is difficult to probe. 

Following Refs.~\cite{Grossman:2025cov,Cheng:2025fak}, we consider the one-loop contribution to the axion mediated force in the presence of the axion DM background generated by the quadratic axion-fermion couplings. Despite the two interaction terms, Eqs.~(\ref{eqn:L2_SI},\ref{eqn:L2_SB}) appearing very similar in structure, they have very different effects in terms of this force.

The basic idea of the background enhancement effect is that the axion DM background will change the vacuum state from $\ket{0}$ to $\ket{n}$, and thus there will be an extra contribution from the occupation number which can modify the axion propagator. The complete propagator, $D_{\rm bk}$, contains both vacuum and background terms~\cite{Grossman:2025cov,Cheng:2025fak,Barbosa:2024pkl}
\begin{equation}
    D_{\rm bk}=\frac{i}{k^2-m_a^2+i\epsilon}+2\pi\delta(k^2-m_a^2)\Theta(k^0)f(\mathbf{k})\, ,
    \label{eqn:background_propagator}
\end{equation}
where $\Theta$ is a Heaviside function and $f(\mathbf{k})$ is the DM momentum distribution function.

The background contribution in Eq.~\eqref{eqn:background_propagator} is
proportional to the on-shell factor $\delta(k^2-m_a^2)$. It therefore does
not contribute to the tree-level exchange potential, where the exchanged
axion carries spacelike momentum set by the momentum transfer between the
sources. By contrast, at one-loop the internal momenta are integrated over,
so an internal axion line can go on-shell and sample the ambient axion
occupation number. The background enhancement of the force is therefore a
loop-level effect. Following the calculation of scattering amplitudes and the two particle interaction potential in Refs.~\cite{Cheng:2025fak,Grossman:2025cov}, one can see that the SI quadratic coupling, Eq.~\eqref{eqn:L2_SI}, receives no background enhancement, while the SB case, Eq.~\eqref{eqn:L2_SB}, leads to the position space potential
\begin{equation}
    V_{\rm bk}(r)=-\frac{\alpha_\psi^2\rho_a}{4\pi rf_a^4m_a^2}e^{-2\kappa_0^2r^2}\, ,
    \label{eqn:enhanced_force}
\end{equation}
where $\kappa_0\simeq m_a v_a/\sqrt{2}$, and $v_a\sim10^{-3}c$ in the local DM halo assuming a Maxwell-Boltzmann distribution. The resulting long-range force is not only background enhanced by $\rho_a$, but is also spin-independent. 

The sign of the effective form factor in Eq.~\eqref{eqn:enhanced_force} is fixed within the de Broglie wavelength and leads to an attractive force. Outside of the de Broglie radius Eq.~\eqref{eqn:enhanced_force} is only approximate, and the sign of the form factor can change due to DM wakes~\cite{VanTilburg:2024xib,Day:2023mkb}. We neglect such effects and consider only the cut-off generated by the Gaussian outside the de Broglie wavelength. This approximation only affects our results for $m_a\gtrsim 10^{-12}\text{ eV}$, where in any case the effects on Hulse-Taylor systems become negligible.

\subsection{Tachyonic Phase Transition for the Light QCD Axion}

For the light QCD axion, the operator Eq.~\eqref{eqn:L2_SB} induces a tachyonic phase transition that contributes to the long-range force in vacuum (i.e. without background enhancement)~\cite{Hook:2017psm}. Consider the axion self-interaction potential energy including the effective mass term introduced by neutron star matter 

\begin{equation}
    V(a)\simeq m_a^2f_a^2\left[1-\cos\left({\frac{a}{f_a}}\right)\right]-\frac{\alpha_\psi}{f_a^2}\frac{Y_{\psi}\,\rho_\textrm{NS}}{m_n}a^2
\end{equation}
which gives an effective mass for the axion
\begin{equation}
    m_{\rm {\rm eff}}^2=m_a^2-\frac{\alpha_\psi}{f_a^2}\frac{Y_{\psi}\,\rho_\textrm{NS}}{m_{n}}\, .
\end{equation}
where $Y_\psi$ is the fermion abundance. The condition for a tachyonic solution happens when the phase transition energy from the axion potential is larger than the gradient energy, which is
\begin{equation}
    |{m_{\rm{\rm eff}}^2}|f_a^2>\frac{f_a^2}{r^2}
\end{equation}
leading to
\begin{equation}
    r^2>\frac{1}{|{m_{\rm{\rm eff}}^2}|}\equiv r_{\rm{crit}}^2
\end{equation}
once $m_{\rm{\rm eff} }^2<0$ and $r_{\rm{crit}}<r_\textrm{NS}$ are both satisfied, the tachyonic phase transition happens~\cite{Hook:2017psm}. The first condition implies a upper bound for $f_a$ , which is $f_a^2<\frac{\alpha_\psi n_\psi}{m_a^2}$, and substituting parameters 
\begin{equation}
f_a<2.5\times10^{19}{\rm{GeV}}\left(\frac{\alpha_\psi}{5.4\,{\rm{MeV}}}\right)^{1/2}\left(\frac{\bar{n}_\psi}{0.15\,\rm{fm}^3}\right)^{1/2}\left(\frac{m_a}{10^{-13}\rm{eV}}\right)^{-1}    
\end{equation}
and
\begin{equation}
    r_{\rm{NS}}>\frac{1}{\sqrt{\frac{\alpha_\psi}{f_a^2}\frac{Y_{\psi}\,\rho_{\rm{NS}}}{m_n}-m_a^2}}\,,
\end{equation}

After the tachyonic phase transition inside the neutron star, an axion profile will be sourced due to the field value difference between the vacuum inside and outside the neutron star which is~\cite{Hook:2017psm}
\begin{equation}
    a(r)=\frac{q_{\rm{\rm eff}}e^{-m_a r}}{r}\, .
\end{equation}

A fully self-consistent treatment of this regime requires solving simultaneously for the scalar profile and the neutron star structure, since scalar sourcing can modify the equation of state, stellar configuration, and associated astrophysical observables~\cite{Balkin:2023xtr,Gomez-Banon:2024oux,Witte:2025ilt}. Here we instead adopt the fixed-background scalar-charge approximation in order to map the binary-timing constraint onto the light-QCD axion parameter space. The resulting tachyonic bounds should therefore be regarded as complementary to constraints based directly on scalar-induced modifications of neutron star structure, cooling, and pulsar evolution. Where $q_{\rm{eff}}=\pm \pi f_ar_{\rm{NS}}$ and the gradients of the axion field source an additional effective potential between two neutron stars 
\begin{equation}
    V_{\rm tach.}(r)=-\frac{q_1q_2e^{-m_a r}}{4\pi r}\, ,
\end{equation}
with $q_{1,2}=\pm 4\pi (\pi f_a r_{\rm{NS}})$ and the sign is randomly chosen after the tachyonic phase transition happens inside neutron stars. Thus, the tachyonic force can be either attractive or repulsive, and the sign in a given system cannot be known unless the presence of the force is detected, i.e. it is unknown when one can only set limits.

\subsection{Axion Radiation from Binary Systems}

We can derive the equation of motion for an axion with the SB quadratic fermion coupling Eq.~\eqref{eqn:L2_SB}
\begin{equation}
    \partial_\mu\partial^\mu a+m^2_a a=\frac{\alpha_\psi}{f_a^2}a\bar{\psi}\psi=\frac{\alpha_\psi}{f_a^2}\frac{\sqrt{\rho_a}}{m_a}\frac{Y_\psi\rho_{\rm{NS}}}{m_n}=\frac{\alpha_\psi}{f_a^2}\frac{\sqrt{\rho_a}}{m_a}Y_\psi N_n\delta^{(3)}(\mathbf{x}-\mathbf{x_0})\, ,
\end{equation}
where $N_n$ is the total neutron number. To derive the energy loss we ignore the slowly oscillating temporal component of the axion background in the regime where the axion mass is much smaller than the binary neutron star orbital frequency. The source term in the action for scalar radiation is~\cite{Huang:2018pbu,Ross:2012fc}
 \begin{equation}
     A_a^{\mathrm{source}}=\int dt\,\mathbf{J}a(t,\mathbf{x})\,.
 \end{equation}
We have to consider the action for a neutron star binary and the coupling to the axion, which gives
\begin{equation}
    A_{a,\,\rm{NS}}^{\rm{source}}=-\sum_{n=1,2}\int d\tau\left(M_n+q_n\frac{2a}{M_{\mathrm{Pl}}}+p_n(\frac{2a}{M_{\mathrm{Pl}}})^2+...\right)\, ,
\end{equation}
where $M_{\rm Pl}=1/\sqrt{8\pi G}$ is the reduced Planck mass. Thus for the background enhanced case we have
\begin{equation}
    \mathbf{J}=\frac{q_n}{M_{\rm{Pl}}}=\frac{\alpha_\psi\sqrt{\rho_a}Y_{\psi}\,N_n}{f_a^2m_a}
\end{equation}
and the effective scalar change for neutron stars in the axion DM  background is
\begin{equation}
 q_{n,\rm{bk}}=\frac{\alpha\sqrt{\rho_a}Y_{\psi}\,N_n M_{\rm{Pl}}}{2f_a^2m_a}\, .
\end{equation}
Similarly for the tachyonic phase transition the induced scalar change is 
\begin{equation}
    q_{n,\rm{tachy.}}=\pm 2\pi^2 f_a r_{\rm{NS}}M_{\rm{Pl}}
\end{equation} 
Then the dipole and quadrupole axion radiation power due to neutron star binary inspiral are~\cite{Hook:2017psm,Huang:2018pbu} 

\begin{align}
    P_a^{l=1}&=\frac{4}{12\pi}\frac{(q_1M_2-q_2M_1)^2}{M^2M_{\mathrm{PI}}^2}\left(1-\frac{m_a^2}{\omega^2}\right)^{3/2}r^2\omega^4\Theta\left(\omega-m_a\right)\,,\\
    P_a^{l=2}&=\frac{16}{15\pi}\frac{(q_1M_2^2+q_2M_1^2)^2}{M^4M_{\mathrm{PI}}^2}\left(1-\frac{m_a^2}{4\omega^2}\right)^{5/2}r^4\omega^6\Theta\left(\omega-\frac{m_a}{2                             }\right)\, ,\
\end{align}
where $M_{1,2}$ are the individual neutron star masses and $M=M_1+M_2$ is the combined mass. In all the formulae we have included the self binding energy of neutron stars, i.e. $m_n N_{n,i}=M_i+\frac{3}{5}\frac{GM_i^2}{R_i}$, and this gives a leading contribution to the dipole emission. For eccentric relativistic binaries, apsidal precession may further modify the harmonic structure of the scalar radiation ~\cite{Xu:2026cky}.

\begin{figure}
    \centering
    \includegraphics[width=1\linewidth]{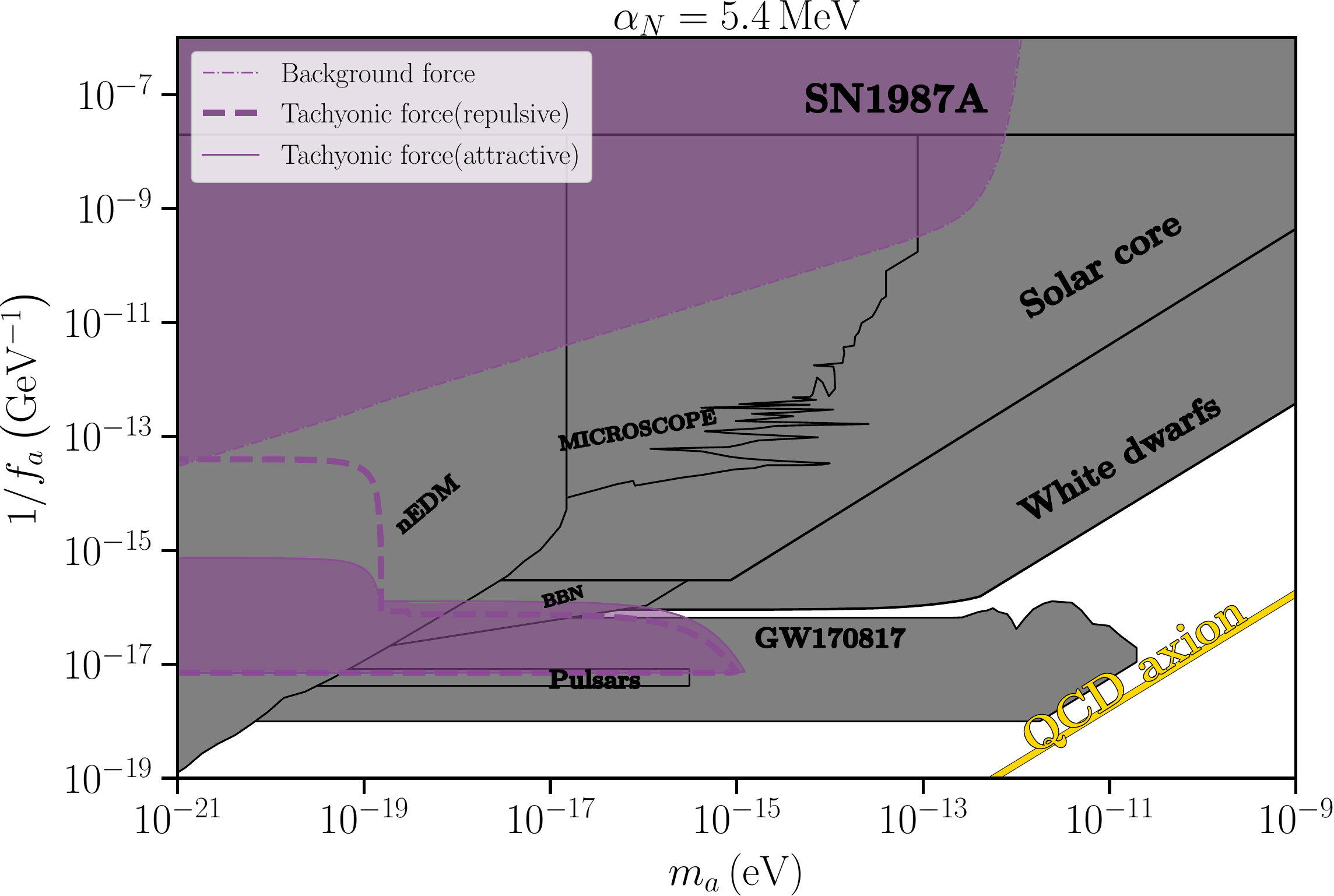}
    \caption{Constraints on the axion-gluon coupling (inverse decay constant) as a function of the axion mass. The purple region shows the parameter space excluded for the ``light QCD axion'' scenario for the benchmark value of the axion-neutron quadratic coupling coefficient, $\alpha_N=5.4\,{\rm MeV}$, taking into account both long-range forces and emission channels. The upper part of the exclusion is for the background enhanced force. The lower region is for the tachyonic phase transiton, which is independent of the DM  background. The tachyonic constraint should be compared with the ``pulsars'' constraint of Ref.~\cite{Hook:2017psm}: our revised bound in this region is more robust and somewhat stronger. The other excluded regions are from axion-DM background dependent constraints: MICROSCOPE~\cite{Gue:2025nxq}, nEDM~\cite{Abel:2017rtm}, Big Bang nucleosynthesis~\cite{Blum:2014vsa}, and background independent constraints from SN1987A~\cite{Springmann:2024ret}, Solar Core~\cite{Hook:2017psm}, white dwarfs~\cite{Balkin:2022qer} and GW170817~\cite{Zhang:2021mks}.}
    \label{fig:neutron_main}
\end{figure}

\section{Results}
\label{sec:results}

\begin{table}[htbp]
    \centering
    \begin{tabular}{lll}
        \hline
        Parameter & Benchmark value & Comment \\
        \hline
        $\alpha_N$
            & $5.4~\mathrm{MeV}$
            & Neutron quadratic-coupling coefficient \\

        $\alpha_\mu$
            & $2.77~\mathrm{eV}$
            & Muon quadratic-coupling coefficient \\

        $\alpha_e$
            & $0.30~\mathrm{eV}$
            & Electron quadratic-coupling coefficient \\

        $Y_n$
            & $1$
            & Neutron abundance, $Y_\psi \equiv N_\psi/N_n$ \\

        $Y_e$
            & $\sim10^{-2}$
            & Electron-to-neutron number ratio \\

        $Y_\mu$
            & $\sim10^{-2}$
            & Muon-to-neutron number ratio \\

        $\rho_a$
            & $0.4~\mathrm{GeV\,cm^{-3}}$
            & Fiducial Milky Way halo axion-DM energy density \\

        $r_{\mathrm{NS}}$
            & $10~\mathrm{km}$
            & Fiducial neutron star radius \\

        $M_1$
            & $1.44\,M_\odot$
            & Mass of the first neutron star \\

        $M_2$
            & $1.39\,M_\odot$
            & Mass of the second neutron star \\
        \hline
    \end{tabular}
    \caption{Benchmark parameters used in the numerical estimates and
    exclusion limits. We define the constituent abundance as
    $Y_\psi\equiv N_\psi/N_n$, where $N_\psi$ is the number of fermions
    of species $\psi$ and $N_n$ is the neutron number. The orbital
    parameters correspond to representative values for the
    Hulse--Taylor binary PSR~B1913+16.}
    \label{tab:benchmark_parameters}
\end{table}

Here we bring forward the expressions from the previous sections and derive constraints, an instructive set of benchmark parameters are shown in Tab.~\ref{tab:benchmark_parameters}. The parametric expression of the ratio between the background enhanced axion potential and Newtonian gravitational potential, $V_N$, can be written as 

\begin{align}
    \frac{V_{\rm{bk}}}{V_N}&=\beta_{\rm{bk}}=\frac{Y_{\psi}^2\,\alpha_\psi^2\rho_a2 M_{\mathrm{PI}}^2}{ f_a^4m_a^2m_n^2} \, \nonumber\\
    &\simeq1.2\times10^{-5}\left(\frac{\alpha_\psi}{5.4\,\mathrm{MeV}}\right)^2\left(\frac{Y_{\psi}}{0.01}\right)^2\left(\frac{10^{13}\mathrm{GeV}}{f_a}\right)^4\left(\frac{10^{-21}\mathrm{eV}}{m_a}\right)^2\left(\frac{\rho_a}{0.4\mathrm{GeV/cm^3}}\right)\, ,
\end{align}
While for the tachyonic force we have
\begin{equation}
    \beta_{\rm{tach}}=\mp\frac{4\pi^3 f_a^2r^2_{\rm{NS}}8\pi M_{\rm{Pl}}^2}{M_1M_2}e^{-m_a r}\simeq\mp 0.187\left(\frac{f_a}{10^{14}{\rm{GeV}}}\right)^2\,,
\end{equation}
where we assume $m_a^{-1}\gg r$ to provide the simplified numerical scaling.

We also have the dipole and quadrupole radiation for both background and tachyonic induced scenarios. For the background enhanced case we have the dipole radiation 
\begin{equation}
    \frac{P_{a,\rm{bk}}^{l=1} f_{\mathrm{dip}}(e)}{P_{GW} f_{\mathrm{qua}}(e)}=\xi^{(1)}_{\mathrm{bk}}\simeq1.36\times 10^{-8}\left(\frac{10^{14}\mathrm{GeV}}{f_a}\right)^4\left(\frac{10^{-22}\mathrm{eV}}{m_a}\right)^2\left(\frac{\alpha_\psi}{5.4\,\mathrm{MeV}}\right)^2\left(\frac{\rho_a}{0.4\,\mathrm{GeV/cm^3}}\right)\, ,
\end{equation}
and quadrupole radiation:
\begin{align}
    \frac{P_{a,\rm{bk}}^{l=2} f_{\mathrm{qua}}(e)}{P_{GW} f_{\mathrm{qua}}(e)}&=\xi^{(2)}_\mathrm{bk}=\frac{\left(q_1\frac{M_2}{M_1}+q_2\frac{M_1}{M_2}\right)^2}{M^2}\nonumber\\
    &\simeq8\pi\left(\frac{M_{\mathrm{PI}}\alpha_\psi\sqrt{\rho_a}}{f_a^2m_am_\psi }\right)^2\left[\frac{M_2}{M}\left(1+\frac{3}{5}\frac{GM_1}{R_1}\right)+\frac{M_1}{M}\left(1+\frac{3}{5}\frac{GM_2}{R_2}\right)\right]^2\nonumber \\
&\simeq2.53\times10^{-8}\left(\frac{10^{14}\mathrm{GeV}}{f_a}\right)^4\left(\frac{10^{-22}\mathrm{eV}}{m_a}\right)^2\left(\frac{\alpha_\psi}{5.4\mathrm{MeV}}\right)^2\left(\frac{\rho_a}{0.4\mathrm{GeV/cm^3}}\right)
\end{align}

In the scaling estimate above we neglected the phase-space factor $\left(1-m_a^2/\omega^2\right)^{3/2}$, which is appropriate in the small-axion-mass limit. For the Hulse--Taylor binary we use the representative parameters
$M_1 \approx 1.44 M_\odot$, $M_2 \approx 1.39 M_\odot$, $e \approx 0.617$, $T \approx 7.75~{\rm hr}$, and the semi-axis $a \approx 1.95 \times 10^6~{\rm km}$. The phase-space suppression becomes important when the axion mass approaches the characteristic orbital frequency~\cite{Mohanty:1994yi,Damour:2014tpa,Weisberg:2016jye}.

For the tachyonic scenario we have the dipole and quadrupole radiation:
\begin{align}
    \frac{P_{a,\rm{tach}}^{l=1} f_{\mathrm{dip}}(e)}{P_{GW} f_{\mathrm{qua}}(e)}&=\xi^{(1)}_{\rm tach}\simeq1.39\times10^{-6}\left(\frac{f_a}{10^{14}\,\rm{GeV}}\right)^2\\
    \frac{P_{a,\rm{tach}}^{l=2} f_{\mathrm{qua}}(e)}{P_{\rm{GW}}f_{\rm{qua}}(e)}&=\xi^{(2)}_{\rm tach}\simeq3.19\times 10^{-8}\left(\frac{f_a}{10^{14}\,\rm{GeV}}\right)^2
\end{align}
We now place constraints on the axion parameters using the observed constraints on $(\beta,\xi)$ derived in Sec.~\ref{sec:gr_tests}.

\begin{figure}[htbp]
    \centering
    \includegraphics[width=1\linewidth]{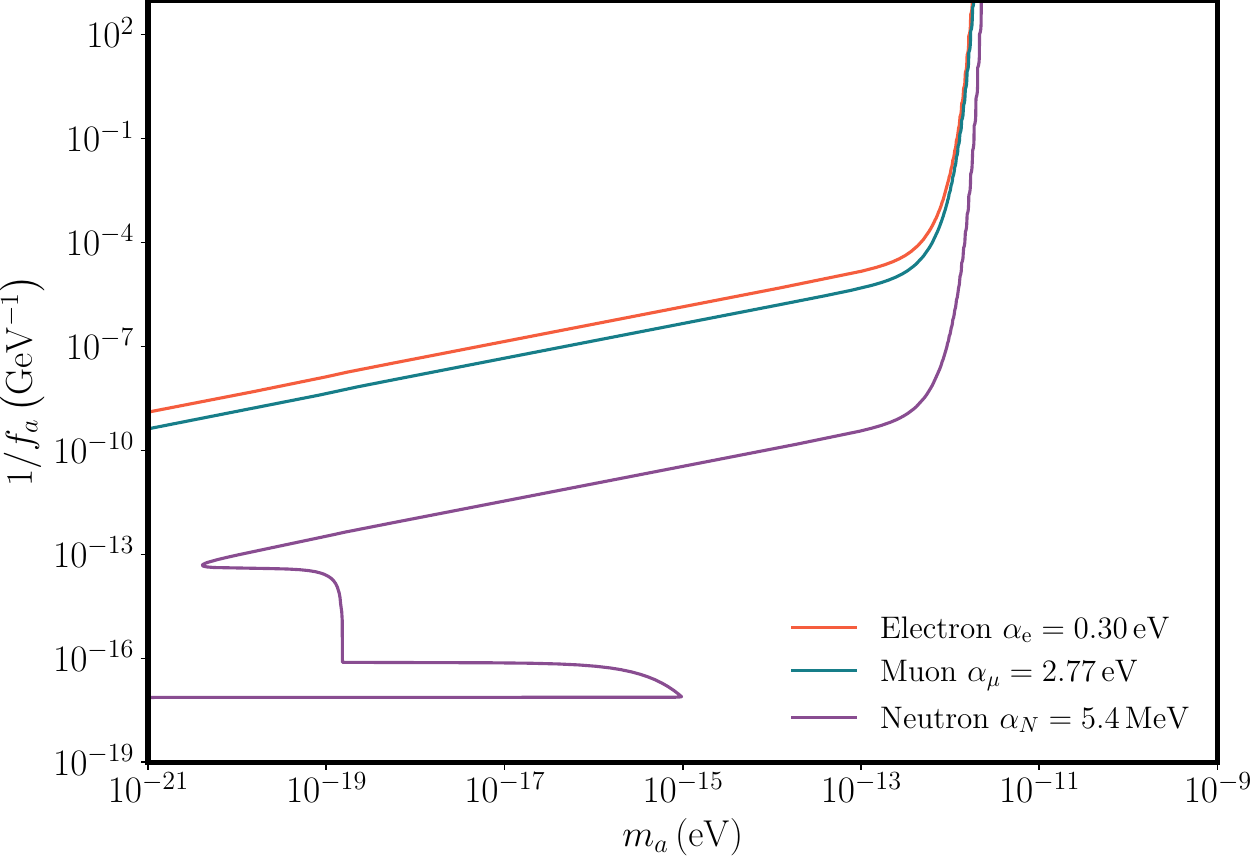}
    \caption{Constraint contours for inverse decay constant as a function of the axion mass in the ``light QCD axion'' model. The different colored lines correspond to different neutron star constituents, with the associated benchmark values of $\alpha_\psi$. The contours in the upper-left region arise from the background enhanced force, while those in the lower left region are due to the attractive tachyonic force. For the reference value of the electron coupling $\alpha_e$, no constraint is derived for the tachyonic force. The neutron coupling gives the strongest limit on $1/f_a$ over the entire parameter space.}
    \label{fig:ma_fa_emuneu}
\end{figure}

Fig.~\ref{fig:neutron_main} shows our constraints on $1/f_a$ in the ``light QCD axion'' scenario for the Hulse-Taylor binary. The constraints from the background mediated force alone are in the top part of the figure at large values of $1/f_a$ (small $f_a$). Our limits exceed those from supernova cooling~\cite{Springmann:2024ret} for $m_a\lesssim 10^{-12}\text{ eV}$. Over the range $10^{-20}\text{ eV}\lesssim m_a\lesssim 10^{-13}\text{ eV}$ our limits are approximately one order of magnitude weaker than the MICROSCOPE fifth force limits~\cite{Gue:2025nxq}, however our results have validity at lower and higher masses where the MICROSCOPE limits are not constraining. Our limits are weaker than those from BBN~\cite{Blum:2014vsa}, but complementary to these due to the distinct cosmological and astrophysical dependencies and uncertainties. Complementary constraints on this light-QCD axion parameter space arise from the backreaction of the sourced axion field on neutron star structure~\cite{Balkin:2023xtr}, isolated-neutron-star cooling~\cite{Gomez-Banon:2024oux}, and the death line of rotation-powered pulsars~\cite{Witte:2025ilt}. These probes exploit different stellar observables and modelling assumptions from the binary-timing analysis considered here, and we therefore do not overlay them in Fig.~\ref{fig:neutron_main}.
Once we include the effect of the tachyonic force, we gain complementary and combined sensitivity in the small $1/f_a$ region. Here, our new limits can be considered as extended and revised version of the ``pulsars'' limit~\cite{Hook:2017psm} (which uses only the requirement that the axion force is always weaker than gravity). In the conservative case, the tachyonic force is attractive (solid lines), and our analysis gives a larger constrained region (stronger constraint) than Ref.~\cite{Hook:2017psm}. If the force is repulsive, however, our limits become stronger still by an order or magnitude extending to larger $1/f_a$. For the tachyonic force, our limits are weaker than those from GW170817~\cite{Zhang:2021mks}.

\begin{figure}
    \centering
    \includegraphics[width=1\linewidth]{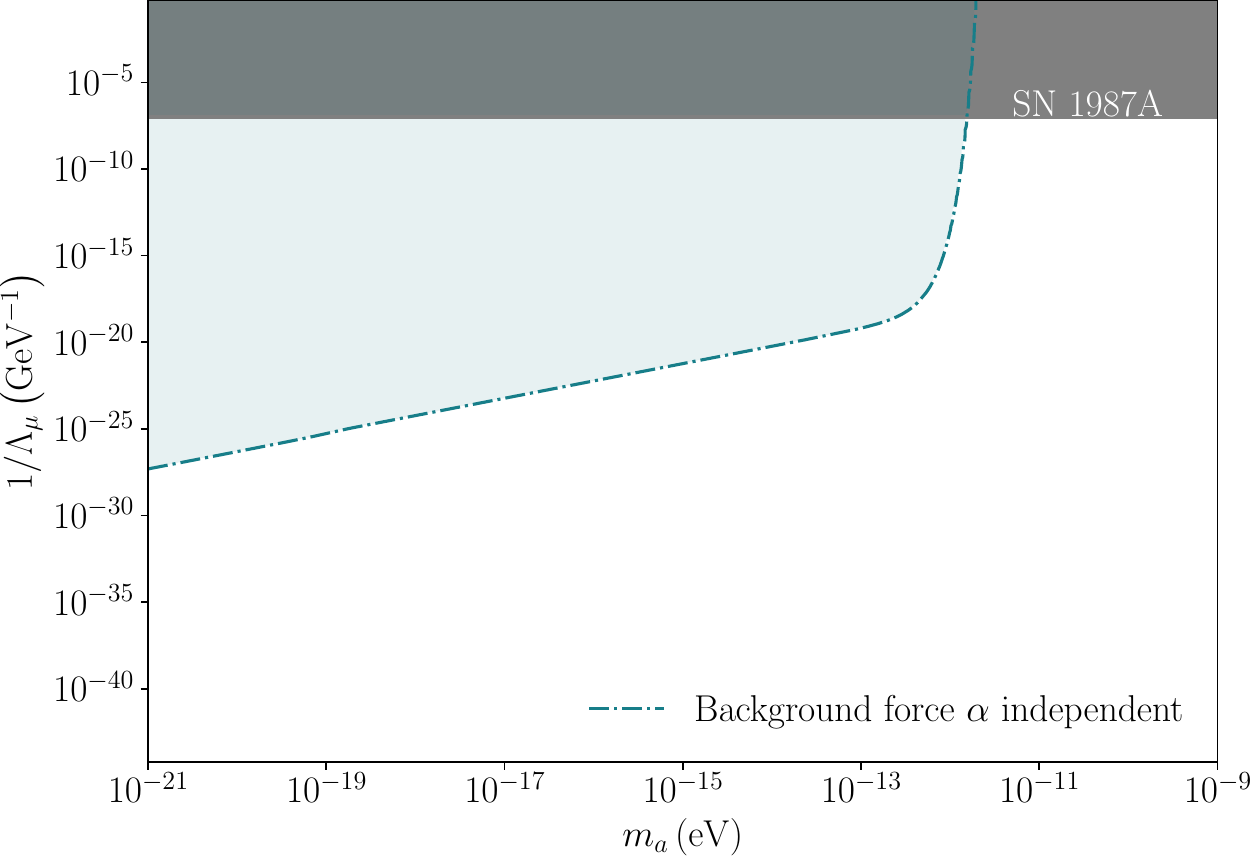}
    \caption{Constraints on the quadratic axion--muon coupling scale $\Lambda_\mu$ as a function of the axion mass. The green shaded region is excluded by the background-enhanced long-range force and the gray shaded region shows the SN1987A cooling constraint derived in Appendix~\ref{sec:mumu_annihilation}. Unlike the electron-coupling case, Fig.~\ref{fig:axion_electron coupling_lambda}, 
    the BBN constraint for this case is irrelevant since muons decouple much earlier than electrons.}
    
    \label{fig:axion_muon coupling_lambda}
\end{figure}
\begin{figure}
    \centering
    \includegraphics[width=1\linewidth]{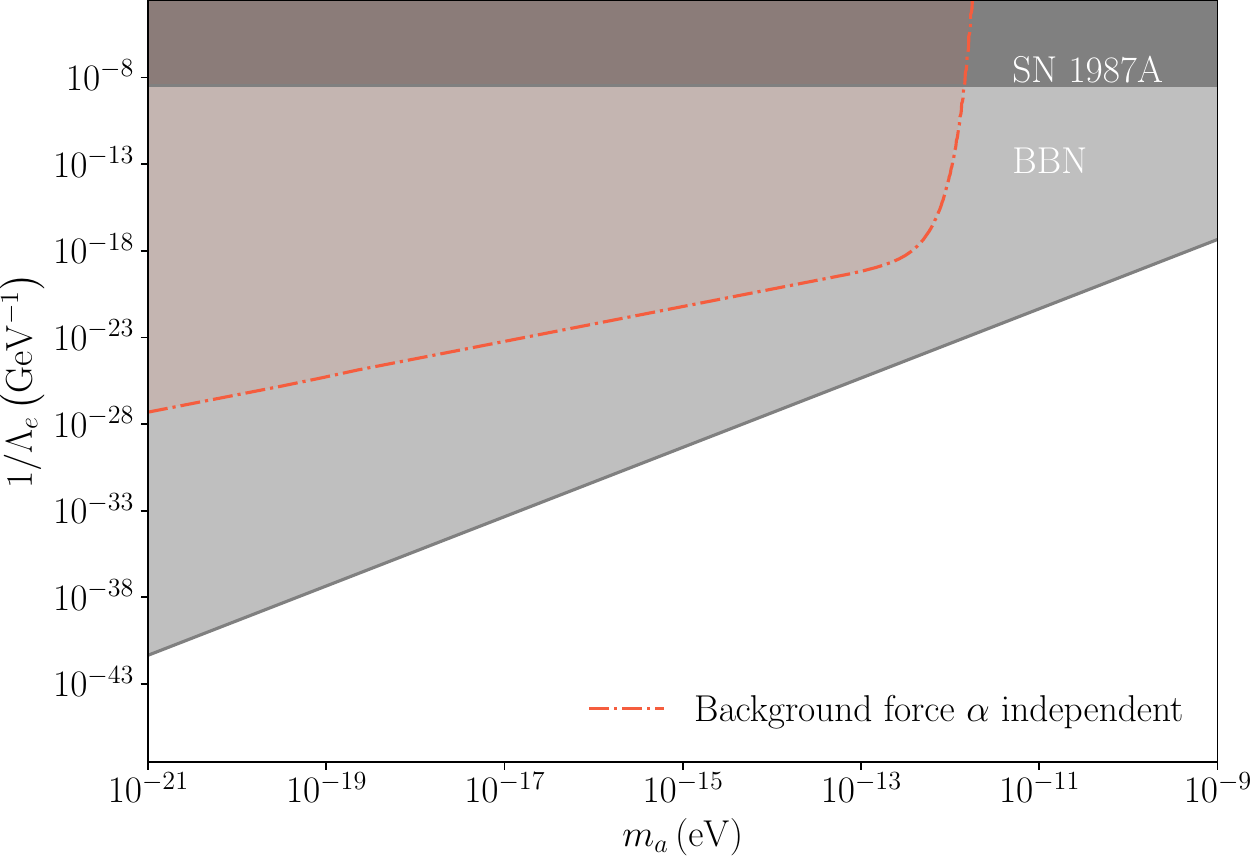}
    \caption{Constraints on the quadratic axion--electron coupling scale $\Lambda_e$ as a function of the axion mass. The red shaded region is excluded by the background-enhanced long-range force. Tachyonic instability provides no additional constraint for the light-QCD benchmark considered above. The BBN and SN1987A limits are from Ref.~\cite{Day:2023mkb}.}
    \label{fig:axion_electron coupling_lambda}
\end{figure}

Fig.~\ref{fig:ma_fa_emuneu} shows constraints on $1/f_a$ comparing those arising from the electron, muon, and neutron couplings using the benchmark values of $\alpha_\psi$. The neutron mediated forces always lead to the strongest limit in the considered parameter space, unsurprisingly due to the dominant neutron composition in the neutron stars and larger value of $\alpha_N$.

Independently of the value of $\alpha_\psi$, we present limits on $\Lambda_\mu$ and $\Lambda_e$ EFT scales in Figs.~\ref{fig:axion_muon coupling_lambda} and \ref{fig:axion_electron coupling_lambda} respectively. For the muon coupling, we derive the relevant supernova energy loss bound in the Appendix. Our bound from the background mediated force exceeds the supernova limit for $m_a\lesssim 10^{-12}\text{ eV}$ and to our knowledge, is the strongest available bound in this mass range for the $\Lambda_\mu$ parameter space. For the electron coupling, the background force limit also exceeds the supernova energy loss limit, but remains weaker than, but complementary to, the limit from BBN throughout the parameter space.
Quadratically coupled ultralight scalars are also constrained by white dwarf mass--radius measurements, with particularly strong sensitivity to electron couplings through modifications of the stellar equation-of-state and equilibrium structure~\cite{Bartnick:2025lbg,Bartnick:2025lbv}. These stellar-structure constraints probe a different physical effect from the binary long-range force and radiation considered here and are therefore complementary to our $\Lambda_e$ bound.

\begin{figure}
    \centering
    \includegraphics[width=1\linewidth]{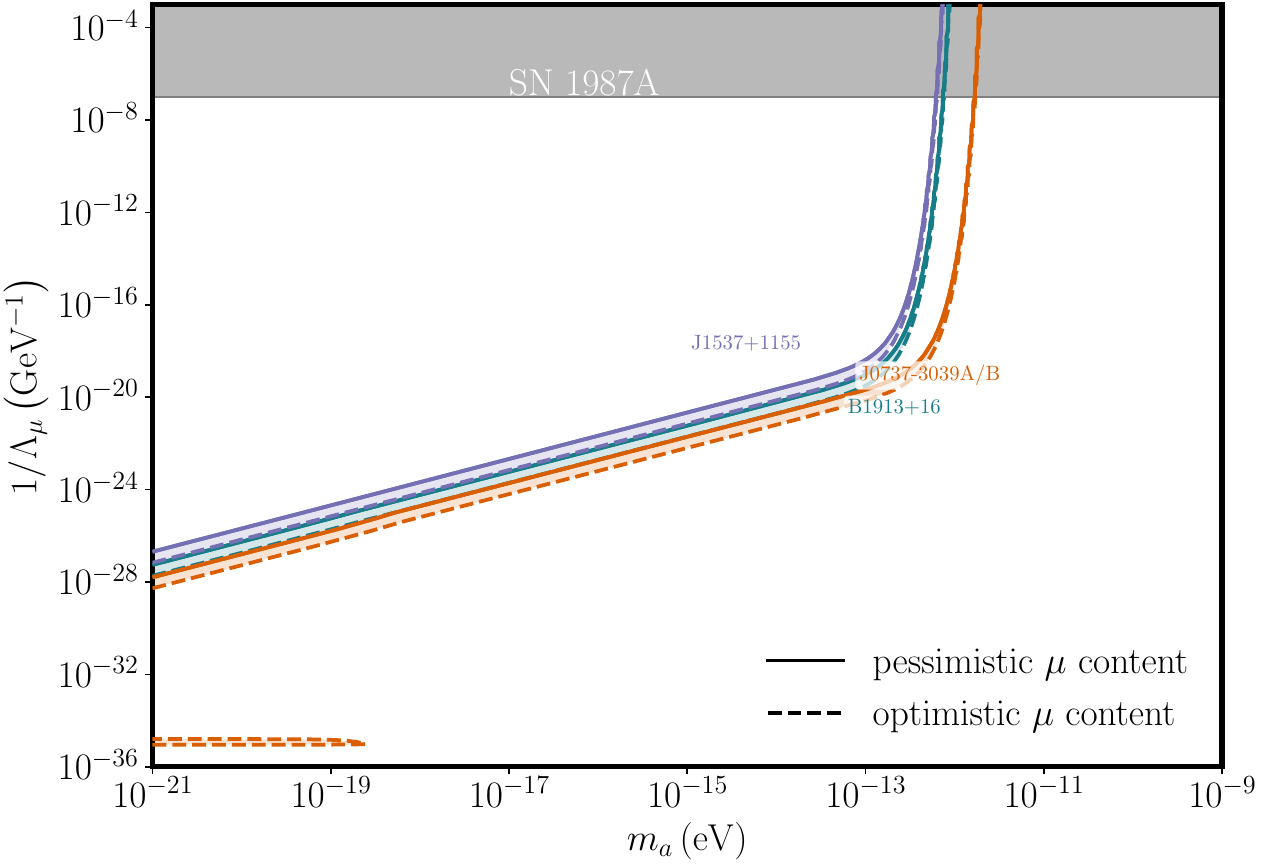}
    \caption{Constraints on the axion-muon quadratic coupling EFT scale for two other binary pulsars, J0737-3039A/B~\cite{Kramer:2021jcw} and J1537+1155~\cite{Ding:2021bhk} (Hulse-Taylor is B1913+16.~\cite{1982ApJ...253..908T}). We model pessimistic and optimistic muon content using the equation-of-state models in Ref.~\cite{Dror:2019uea} taking a reference neutron star mass of 1.3 M$_\odot$. The lower orange shaded region is excluded by tachyonic-instability-induced long-range forces between the neutron stars in J0737-3039A/B system.}
    \label{fig:systematics}
\end{figure}

Finally, we extend our analysis to include other binary pulsars, and to consider the effect of uncertainties in ths neutron star equation-of-state. We consider both of these effects for the muon coupling EFT parameter, $\Lambda_\mu$, in Fig.~\ref{fig:systematics}, following Ref.~\cite{Dror:2019uea}. We consider two other binary systems, J0737-3039A/B~\cite{Kramer:2021jcw} and J1537+1155~\cite{Ding:2021bhk} (Hulse-Taylor is B1913+16). J0737-3039A/B gives the strongest constraint everywhere in parameter space, but is less well studied overall than the historic Hulse-Taylor system. To estimate the uncertainties originating from the unknown equation-of-state, we leverage the fact that for all of these binary systems the neutron star mass is close to 1.3 M$_\odot$. In this range of neutron star mass, Fig. 3 in Ref.~\cite{Dror:2019uea} shows an uncertainty on the muon content between 1\% (pessimistic) and 3\% (optimistic) depending on the chosen equation-of-state. This uncertainty is reflected in the shaded region for each system in Fig.~\ref{fig:systematics}. The uncertainty on the muon content has little effect on our constraints for the background mediated force. For the tachyonic force, however, we see that Hulse-Taylor itself only gives a limit in the case of an optimistic muon content.

\section{Conclusions}
\label{sec:conclusions}

The Hulse--Taylor binary pulsar provides a clean laboratory for testing new long-range interactions and additional radiation channels, because its orbital evolution is measured precisely and agrees closely with the predictions of general relativity. In this work we used this system to constrain quadratic axion couplings to stable neutron star constituents. Such operators are well motivated by shift symmetry breaking effects and can generate spin-independent forces between compact objects, especially in the presence of an ambient axion-DM background. The same interactions also source axion radiation from binary motion, providing a second contribution to the orbital-period decay.

We formulated the binary dynamics in terms of two phenomenological parameters: $\beta$, which captures modifications to the conservative force, and $\xi$, which captures the fractional enhancement of radiative energy loss. Requiring the modified orbital decay to remain consistent with the observed Hulse-Taylor period derivative gives a direct constraint in the $\beta$-$\xi$ plane. We then mapped this constraint onto axion parameter space for quadratic couplings to neutrons, electrons, and muons, including both the background enhanced long-range force and the force induced by a tachyonic axion profile inside neutron stars.

For axion-neutron couplings, the resulting bounds can be interpreted as constraints on light QCD axion scenarios, where the quadratic coupling is tied to the axion-gluon interaction. For leptonic couplings, neutron star electron and muon populations allow the Hulse-Taylor system to probe axion-electron and axion-muon quadratic interactions. In the muon case, the absence of a BBN bound makes binary-pulsar dynamics a particularly useful complementary probe, while the SN1987A cooling estimate derived in Appendix~\ref{sec:mumu_annihilation} is subdominant for the parameter space considered here.

Our results demonstrate that precision binary-pulsar timing can provide robust and independent constraints on quadratic axion-fermion couplings. This framework can be extended to other relativistic binaries, to systems with different neutron star equations-of-state, and to more detailed treatments of axion radiation and environmental axion-DM effects. 

\section*{Acknowledgements}

We thank Sebastian Ellis, Jim Halverson,  and Bingrong Yu for helpful discussions. DJEM is supported by an Ernest Rutherford Fellowship (Grant No. ST/T004037/1).  SB, MF and DJEM were supported by a consolidator grant (Grant No. ST/X000753/1) from the Science and Technologies Facilities Council (STFC), United Kingdom.  ZY acknowledges the support of the China Scholarship Council program
(Project ID:202406230357) and hospitality of King's College London, where this project was initiated. This paper used open source software \textsc{matplotlib}~\cite{2007CSE.....9...90H}, \textsc{numpy}~\cite{2020Natur.585..357H}, and \textsc{scipy}~\cite{2020NatMe..17..261V}.

\appendix

\section{Binary timing formulae with a modification to GR}
\label{sec:HTGR}
In this appendix we collect the timing formulae used in the
$\chi^2$ analysis of Sec.~\ref{sec:gr_tests}. We parameterise the
additional conservative interaction by
\[
V(r)=-\frac{G(1+\beta)M_1M_2}{r},
\qquad
G_{\rm eff}=G(1+\beta),
\]
The relative orbit therefore obeys the modified Kepler law
\[
n^2a^3=G_{\rm eff}M,
\qquad
n=\frac{2\pi}{P_b},
\qquad
M=M_1+M_2.
\]
The additional radiative energy loss is written
\[
\dot E_{\rm rad}=(1+\xi)\dot E_{\rm GW}^{\rm GR}.
\]
\subsection{Einstein delay}

The Einstein delay contains a gravitational-redshift contribution and a
second-order Doppler contribution,
\[
\Delta_E(t)
=
\int
\left(
\frac{v_1^2}{2c^2}
+
\frac{GM_2}{rc^2}
\right)dt .
\]
The non-metric force modifies the orbital motion, but not the metric
redshift term. Using
\[
v_1=\frac{M_2}{M}v_{\rm rel},
\qquad
v_{\rm rel}^2
=
G_{\rm eff}M
\left(
\frac{2}{r}-\frac{1}{a}
\right),
\]
and retaining only the periodic part of the delay gives
\[
\gamma
=
\frac{e}{anc^2}
\left(
GM_2+\frac{G_{\rm eff}M_2^2}{M}
\right).
\]
Since
\[
\frac{1}{an}
=
(G_{\rm eff}M)^{-1/3}
\left(\frac{P_b}{2\pi}\right)^{1/3},
\]
we obtain
\[
\gamma_{\rm th}
=
e
\left(\frac{P_b}{2\pi}\right)^{1/3}
\frac{(GM)^{2/3}}{c^2}
\left[
\frac{M_2}{M}(1+\beta)^{-1/3}
+
\frac{M_2^2}{M^2}(1+\beta)^{2/3}
\right].
\]

\subsection{Periastron advance}

The relativistic periastron advance is generated by the metric
gravitational interaction, while the semi-major axis inferred from the
observed period is set by the modified Kepler law. Thus
\[
\Delta\omega
=
\frac{6\pi GM}{a(1-e^2)c^2},
\qquad
a^3
=
\frac{G_{\rm eff}MP_b^2}{4\pi^2}.
\]
The periastron-advance rate is therefore
\[
\dot\omega_{\rm th}
=
\frac{6\pi}{(1-e^2)c^2P_b^{5/3}}
\left(2\pi GM\right)^{2/3}
(1+\beta)^{-1/3}.
\]

\subsection{Orbital-period derivative}

We assume that the gravitational radiation is described by the usual GR
quadrupole formula, evaluated on the modified Keplerian orbit, and that
any additional axion radiation contributes a fractional correction
$\xi$ to the total energy loss. The orbital binding energy is
\[
E
=
-\frac{G_{\rm eff}\mu M}{2a}
=
-\frac{1}{2}
\mu
(G_{\rm eff}M)^{2/3}
\left(\frac{2\pi}{P_b}\right)^{2/3},
\]
where
\[
\mu=\frac{M_1M_2}{M}.
\]
Combining $dE/dt=(dE/dP_b)\dot P_b$ with the quadrupole luminosity gives
\[
\dot P_{b,{\rm th}}
=
-\frac{96}{5}
\frac{G^{5/3}}{c^5}
\mu M^{2/3}
(1+\xi)(1+\beta)^{2/3}
\frac{(2\pi)^{8/3}}{P_b^{5/3}}
f(e),
\]
with the Peters--Mathews eccentricity enhancement
\[
f(e)
=
\frac{
1+\frac{73}{24}e^2+\frac{37}{96}e^4
}{
(1-e^2)^{7/2}
}.
\]
\section{Supernova Cooling Bounds for Axion-Muon and Electron Quadratic Coupling}
\label{sec:mumu_annihilation}

Supernova muons have recently been used to constrain light bosons with linear muon couplings, including muon-coupled axions and axion-like particles~\cite{Croon:2020lrf}. To our knowledge, however, there is no dedicated supernova cooling bound in the literature for the axion-muon quadratic operator considered here. We therefore estimate this constraint directly, following the standard supernova energy loss formalism~\cite{Olive:2007aj,Raffelt:1999tx} and using recent studies of muon populations in proto neutron star cores~\cite{Bollig:2017lki,Fischer:2020vie}.

The dominant supernova cooling channel induced by the operator
\begin{equation}
\mathcal{L}_{\rm int}
=
\frac{1}{\Lambda_\mu}a^2\bar\mu\mu
\label{eq:effective_operator}
\end{equation}
is the annihilation process
\begin{equation}
\mu^+ + \mu^- \rightarrow a + a .
\end{equation}

Following the thermal energy-loss formalism developed in
Ref.~\cite{Olive:2007aj}, the energy-loss rate per unit volume can be estimated from Eq.~(3.10) of Ref.~\cite{Olive:2007aj},
\begin{equation}
\Gamma_{\rm ann}
\simeq
n_{\mu^-}n_{\mu^+}
\langle \sigma v\rangle
\langle E_{\rm loss}\rangle,
\label{eq:Gamma_estimate}
\end{equation}
where $\langle E_{\rm loss}\rangle$ denotes the average emitted energy per annihilation event. In the non-relativistic limit relevant for supernova muons,
\begin{equation}
\langle E_{\rm loss}\rangle \simeq 2m_\mu .
\end{equation}

The tree-level matrix element is
\begin{equation}
i\mathcal M
=
\frac{i}{\Lambda_\mu}\bar v(p_2)u(p_1).
\end{equation}
After summing over spins, we obtain
\begin{align}
\sum |\mathcal M|^2
&=
\frac{1}{\Lambda_\mu^2}
{\rm Tr}
\left[
(\slashed p_1+m_\mu)
(\slashed p_2-m_\mu)
\right]
\nonumber\\
&=
\frac{4}{\Lambda_\mu^2}
(p_1\cdot p_2-m_\mu^2)
\nonumber\\
&=
\frac{2}{\Lambda_\mu^2}
(s-4m_\mu^2),
\end{align}
where $s=(p_1+p_2)^2$.

The corresponding annihilation cross section is
\begin{equation}
\sigma_{\mu^+\mu^-\to aa}
=
\frac{1}{8\pi\Lambda_\mu^2}
\sqrt{1-\frac{4m_\mu^2}{s}}.
\label{eq:annihilation_cross_section}
\end{equation}

Inside the supernova core, muons are expected to be mildly non-relativistic,
$T\lesssim m_\mu$, such that
\begin{equation}
s\simeq 4m_\mu^2+m_\mu^2v^2 .
\end{equation}
Expanding Eq.~\eqref{eq:annihilation_cross_section} then gives
\begin{equation}
\sigma v
\simeq
\frac{1}{16\pi\Lambda_\mu^2}v^2,
\end{equation}
indicating that the annihilation process is $p$-wave suppressed.
Using the thermal average
\begin{equation}
\langle v^2\rangle
=
\frac{6T}{m_\mu},
\end{equation}
we obtain
\begin{equation}
\langle \sigma v\rangle
\simeq
\frac{3}{8\pi\Lambda_\mu^2}
\frac{T}{m_\mu}.
\label{eq:sigmav_average}
\end{equation}
The muon number density in the non-relativistic Maxwell-Boltzmann regime is
\begin{equation}
n_\mu
=
2
\left(
\frac{m_\mu T}{2\pi}
\right)^{3/2}
e^{-(m_\mu-\mu_\mu)/T},
\label{eq:muon_density}
\end{equation}
where $\mu_\mu$ is the muon chemical potential.
Recent supernova simulations indicate that sizable muon abundances can develop in the proto-neutron-star core, with
\begin{equation}
\mu_\mu \sim {\cal O}(10\text{--}100)\ {\rm MeV},
\end{equation}
depending on the post-bounce evolution and the equation-of-state
\cite{Bollig:2017lki,Fischer:2020vie}.
Consequently, the Boltzmann suppression factor can be substantially reduced compared to the naive estimate with vanishing chemical potential.

Substituting Eqs.~\eqref{eq:sigmav_average} and \eqref{eq:muon_density} into Eq.~\eqref{eq:Gamma_estimate}, we obtain
\begin{equation}
\Gamma_{\rm ann}({\mu^+\mu^-\to aa})
\simeq
\frac{3}{\pi\Lambda_\mu^2}
\left(
\frac{m_\mu T}{2\pi}
\right)^3
T
\,
e^{-2(m_\mu-\mu_\mu)/T}.
\end{equation}
Equivalently,
\begin{equation}
\Gamma_{\rm ann}({\mu^+\mu^-\to aa})
\simeq
\frac{3}{8\pi^4\Lambda_\mu^2}
m_\mu^3T^4
e^{-2(m_\mu-\mu_\mu)/T}.
\label{eq:Gamma_ann_final}
\end{equation}

The parametric structure of Eq.~\eqref{eq:Gamma_ann_final} is analogous to the thermal annihilation emissivities discussed in Ref.~\cite{Day:2023mkb}, while the overall treatment of the supernova energy-loss rate follows Ref.~\cite{Olive:2007aj}.

Finally, one may derive an approximate supernova cooling bound on $\Lambda_\mu$ using the Raffelt criterion~\cite{Raffelt:1999tx},
\begin{equation}
\frac{\Gamma_{\rm ann}}{\rho}
\lesssim
10^{19}\ {\rm erg}\ {\rm g}^{-1}\ {\rm s}^{-1},
\label{eq:Raffelt_criterion}
\end{equation}
where $\rho$ is the supernova core density.
For typical core conditions,
\begin{equation}
T\sim 30~{\rm MeV},
\qquad
\rho\sim 3\times10^{14}\ {\rm g/cm^3},
\end{equation}
together with
\begin{equation}
\mu_\mu \sim 50~{\rm MeV},
\end{equation}
one obtains an order-of-magnitude lower bound on the effective scale,
\begin{equation}
\Lambda_\mu
\gtrsim
{\cal O}(10^{6}\text{--}10^{7})\ {\rm GeV},
\end{equation}
up to order-one uncertainties associated with the supernova profile, muon abundance, and phase-space treatment. Thus the supernova bound for the axion-muon quadratic coupling is negligible comparing with our bounds from binary pulsar evolution.

For the supernova bound on the axion-electron quadratic coupling, we can follow Ref.~\cite{Day:2023mkb} which gives the relativistic electron-positron annihilation cooling rate
\begin{equation}
    \Gamma_{\rm{ann}}({e^+e^-\to aa})=\frac{7\zeta(3)T^7}{80\pi\Lambda_e^2}\, ,
\end{equation}
which leads to
\begin{equation}
    \Lambda_e\gtrsim 2.8\times10^8\rm{GeV}\, .
\end{equation}

\bibliographystyle{JHEP}
\bibliography{references.bib}

\end{document}